\documentclass[sn-mathphys]{sn-jnl}

\newcommand{\blb}{\boldsymbol{\lambda}}
\newcommand{\bkp}{\boldsymbol{\kappa}}

\begin{document}

\title[Low-frequency magnetic oscillations]{Low-frequency magnetic oscillations\\ induced by strongly electron correlations}

\author{\fnm{Alexei} \sur{Sherman}}\email{alekseis@ut.ee}

\affil{\orgdiv{Institute of Physics}, \orgname{University of Tartu}, \orgaddress{\street{W. Ostwaldi Str 1}, \city{Tartu}, \postcode{50411}, \country{Estonia}}}

\abstract{To explain the low frequencies of quantum oscillations observed in lightly doped cuprates, we consider the two-dimension Hubbard model supplemented with the perpendicular magnetic field. For large Hubbard repulsions, the electron spectrum is investigated using the cluster perturbation theory. Obtained frequencies of magnetic oscillations at small deviations from half-filling are close to those observed experimentally, $F\approx500$~T. They stem from small Fermi surface pockets located in the nodal regions of the Brillouin zone. The pockets are formed by Fermi arcs and less intensive segments, which make the pockets nearly circular.}

\keywords{Magnetic oscillations, Cluster perturbation theory, Hubbard model, Strong correlations}

\maketitle

\section{Introduction}
Experiments with strong magnetic fields $B\gtrsim22$~T, which weaken superconductivity in underdoped yttrium cuprates, reveal properties of their normal state by means of magnetic oscillations (as a review, see \cite{Sebastian}). The observed quantum oscillation frequencies appear to be anomalously small, $F\approx500$~T. In accordance with the Onsager-Lifshitz-Kosevich theory \cite{Shoenberg}, such frequencies correspond to a Fermi surface enclosing approximately 2\% of the Brillouin zone \cite{Doiron}. Experimental results point that the quantum oscillations are connected with Fermi surface pockets located near the nodal points $[\pm\pi/(2a),\pm\pi/(2b)]$ of the Brillouin zone \cite{Sebastian}. Here $a$ and $b$ are the CuO$_2$ plane lattice constants. Angle-resolved photoemission experiments carried out at higher temperatures show that the density of states at the Fermi level of underdoped cuprates is concentrated at Fermi arcs located near the nodal points \cite{Damascelli}. The arcs are disconnected and photoemission data cannot validate the supposition that they are intensive parts of small closed Fermi surface pockets, which needs to explain the quantum-oscillation experiments.

Various states with broken translation symmetry were suggested to explain the appearance of such pockets due to a reconstruction of the electron spectrum (see, e.g., \cite{Millis,Chen,Galitski}). To our knowledge, there are no confirmations of such reconstructions from the photoemission data. Besides, it is unclear how the essential property of underdoped cuprates -- strong electron correlations -- is involved in this symmetry reduction. In Refs.~\cite{Melikyan,Pereg} decreased quantum oscillation frequencies were related to superconducting fluctuations. Fields $22-101$~T used in experiments \cite{Sebastian2} with YBa$_2$Cu$_3$O$_{6+\delta}$ exceeded the upper critical field $H_{c2}$ \cite{Grissonnanche}. Therefore, the use of such a mechanism for interpreting these experiments is hardly feasible. Few known to us works dealing with strongly correlated electrons in high magnetic fields use the dynamic mean-field approximation \cite{Acheche,Markov,Vucicevic}. The case of low doping was not considered in these investigations. Obtained frequencies of magnetic oscillations were an order of magnitude larger than those observed in lightly doped cuprates. Such frequencies correspond to large Fermi surfaces. Small oscillation frequencies, which are comparable to experimental ones, were obtained in Ref.~\cite{Sherman15} applying the strong coupling diagram technique (SCDT). In this work, a simplified description of the irreducible part based on the Hubbard-I approximation was used. In this approximation, small Fermi-surface pockets appear in corners of the Brillouin zone rather than near nodal points.

In this work, we consider electrons described by the two-dimensional (2D) Hubbard model and immersed in a perpendicular quantizing magnetic field. Properties of this system are investigated using the cluster perturbation theory (CPT) \cite{Senechal00,Senechal02}. It is shown that at the electron concentrations $x\approx0.96$, this approach can describe small Fermi-surface pockets near the nodal points. They are responsible for the magnetic oscillations of the density of states (DOS) at the Fermi level with the frequency $F\approx500$~T. It is close to the quantum oscillation frequency observed experimentally \cite{Sebastian}. However, our hole concentration is less than half of the value evaluated in the experimental samples. This discrepancy may be related to the simplicity of the used model and the method of estimating the mobile-electron concentration in samples. The obtained hole pockets contain Fermi arcs supplemented with less intensive parts producing nearly circular contours.

The paper is organized as follows. In the next section, the CPT is discussed, and its main equations are given. In Sec.~3, the magnetic field is added to the consideration. Obtained results on quantum oscillations and shapes of Fermi surfaces are considered in Sec.~4. The last section is devoted to concluding remarks.

\section{Cluster perturbation theory}
In this work, we consider the one-band repulsive Hubbard model on a 2D square lattice. The model is described by the Hamiltonian
\begin{equation}\label{Hubbard}
H=\sum_{\bf ll'\sigma}t_{\bf ll'}a^\dagger_{\bf l'\sigma}a_{\bf l\sigma}+\frac{U}{2} \sum_{\bf l\sigma}n_{\bf l\sigma}n_{\bf l,-\sigma},
\end{equation}
where $a^\dagger_{\bf l\sigma}$ and $a_{\bf l\sigma}$ are electron creation and annihilation operators on the site {\bf l} with the spin projection $\sigma$, $t_{\bf ll'}$ is the hopping integral, $U$ is the Hubbard repulsion, and the electron number operator $n_{\bf l\sigma}=a^\dagger_{\bf l\sigma}a_{\bf l\sigma}$. Below, $t_{\bf ll'}$ is supposed to be nonzero between nearest neighbor sites only, $t_{\bf ll'}=-t\sum_{\bf a}\delta_{\bf l',l-a}$, where ${\bf a}=(\pm a,0),(0,\pm a)$ are four vectors connecting nearest neighbor sites, and $a$ is the lattice spacing.

In CPT \cite{Senechal00,Senechal02}, the lattice is divided into clusters and the Hamiltonian (\ref{Hubbard}) is rewritten as
\begin{equation}\label{CPT}
H=\sum_{\bf L}H_{\bf L}+\sum_{\bf L'\blb'}\sum_{\bf L\blb\sigma}t_{\bf L+\blb,L'+\blb'} a^\dagger_{\bf L'+\blb',\sigma}a_{\bf L+\blb,\sigma}=H_0+H_i,
\end{equation}
where $H_{\bf L}$ is the Hamiltonian (\ref{Hubbard}) written for the cluster labelled by the index {\bf L}, and the second term $H_i$ in (\ref{CPT}) describes electron transfer between neighboring clusters. The site index $\blb$ labels sites inside a cluster such that ${\bf l=L}+\blb$. In CPT, the kinetic-energy term $H_i$ is considered as a perturbation, and Green's functions are calculated using series expansions in powers of this term. In this regard, the CPT is a cluster version of the SCDT \cite{Hubbard,Zaitsev,Vladimir,Metzner,Pairault,Sherman18}. The terms of this expansion are products of hopping integrals $t_{\bf L+\blb,L'+\blb'}$ between neighboring clusters and cluster cumulants of different orders. Two of them, of the first and second orders, are expressed through thermodynamic averages as follows \cite{Kubo}:
\begin{eqnarray}
&&C^{(1)}_{\blb'\blb}(\tau',\tau)=\big\langle{\cal T}\bar{a}_{\blb'\sigma}(\tau')a_{\blb \sigma}(\tau)\big\rangle_c,\label{C1}\\
&&C^{(2)}_{\blb_1\blb_2\blb_3\blb_4}(\tau_1,\sigma_1;\tau_2,\sigma_2;\tau_3,\sigma_3; \tau_4,\sigma_4)\nonumber\\
&&\quad=\big\langle{\cal T}\bar{a}_{\blb_1\sigma_1}(\tau_1)a_{\blb_2\sigma_2}(\tau_2) \bar{a}_{\blb_3\sigma_3}(\tau_3)a_{\blb_4\sigma_4}(\tau_4)\big\rangle_c\nonumber\\
&&\quad\quad-\big\langle{\cal T}\bar{a}_{\blb_1\sigma_1}(\tau_1)a_{\blb_2 \sigma_2}(\tau_2)\big\rangle_c\big\langle{\cal T}\bar{a}_{\blb_3\sigma_3}(\tau_3)a_{\blb_4 \sigma_4}(\tau_4)\big\rangle_c\nonumber\\
&&\quad\quad+\big\langle{\cal T}\bar{a}_{\blb_1\sigma_1}(\tau_1)a_{\blb_4 \sigma_4}(\tau_4)\big\rangle_c\big\langle{\cal T}\bar{a}_{\blb_3\sigma_3}(\tau_3)a_{\blb_2\sigma_2}(\tau_2)\big\rangle_c,\label{C2}
\end{eqnarray}
where ${\cal T}$ is the chronological operator, angle brackets denote the thermodynamic averaging, the subscript $c$ near them indicates that the time dependencies of the operators and the average are determined by the cluster Hamiltonian
\begin{equation}\label{cluster}
{\cal H}_{\bf L}=H_{\bf L}-\mu\sum_{\blb\sigma} a^\dagger_{\blb\sigma}a_{\blb\sigma}
\end{equation}
with the chemical potential $\mu$. If the crystal is divided into identical clusters, cumulants do not depend on the index {\bf L} due to the translation invariance. Therefore, it is omitted in Eqs.~(\ref{C1}) and (\ref{C2}).

\begin{figure}[t]
\centerline{\resizebox{0.99\columnwidth}{!}{\includegraphics{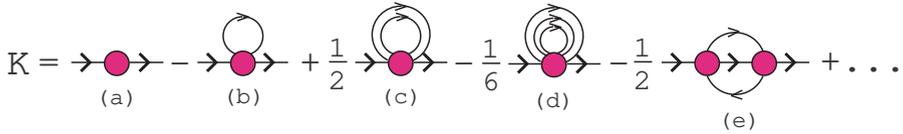}}}
\caption{Diagrams of the first four orders in {\bf K}.} \label{Fig1}
\end{figure}

Each term of the series expansion can be depicted by a diagram, in which arrowed lines stand for hopping integrals and circles for cumulants with the numbers of incoming and outgoing lines equalling to their orders. As in the weak coupling diagram technique with the serial expansion over an interaction, all these diagrams can be classed into reducible and irreducible. For the one-electron Green's function considered in this work, $G_{\blb'\blb}({\bf L'\tau',L\tau})=\langle{\cal T}\bar{a}_{\bf L'+\blb',\sigma}(\tau')a_{\bf L+\blb,\sigma}(\tau)\rangle$, the irreducible diagrams differ from the reducible ones in that the former cannot be divided into two disconnected parts by cutting a hopping line. The sum of all such diagrams is termed the irreducible part {\bf K}. With the use of this quantity, the Green's function can be written as
\begin{equation}\label{Larkin}
{\bf G}({\bf q},j)=[{\bf 1}-{\bf K}({\bf q},j){\bf t}({\bf q})]^{-1}{\bf K}({\bf q},j).
\end{equation}
In this equation, we performed the Fourier transformation over the imaginary times and indices of clusters {\bf L}. The division of the crystal into clusters with the number of sites $n>1$ reduces the translation symmetry of the problem. For identical clusters, this symmetry is retained for translations in the superlattice of clusters only. This leads to a smaller Brillouin zone and momenta ${\bf q}$ belong to this zone. The remaining coordinates of sites in a cluster $\blb$ play the role of matrix indices of the matrices {\bf G}, {\bf K}, and {\bf t} in Eq.~(\ref{Larkin}). In this equation, {\bf 1} is the $n\times n$ unit matrix. The integer $j$ defines the Matsubara frequency $\omega_j=(2j-1)\pi T$ with the temperature $T$. Several lowest order diagrams contributing to {\bf K} are shown in Fig.~\ref{Fig1}.

For clusters containing several lattice sites, the second and higher-order cumulants are rather complicated, and their storage is memory-consuming. Therefore, the CPT approximates the irreducible part {\bf K} by diagram (a) in Fig.~\ref{Fig1} only. This diagram contains the first-order cumulant ${\bf C}^{(1)}$ (\ref{C1}). Comparing this approximation with the SCDT using one-site cumulants and more complicated diagrams, one can see that ${\bf C}^{(1)}$ of the CPT contains several diagrams of the SCDT, which describe electron interactions with spin and charge fluctuations limited by the cluster size. Despite this shortcoming of the CPT, it has the advantage of carrying out calculations at zero temperature and in real frequencies. Since the magnetic-oscillation experiments are performed at extremely low temperatures, these advantages are crucial for the aims of the present work.

In terms of the eigenvectors $\mid\!i\rangle$, $\mid\!m\rangle$ and eigenvalues $E_i$, $E_m$ of the cluster Hamiltonian $H_{\bf L}$, the first-order cumulant reads
\begin{equation}\label{cumulant}
C^{(1)}_{\blb'\blb}(j)=\frac{1}{Z}\sum_{im}\frac{{\rm e}^{-(E_m-\mu p_m)\beta}+{\rm e}^{-(E_i-\mu p_i)\beta}}{{\rm i}\omega_j+E_i-E_m+\mu}\langle i\!\mid\!a_{\blb\sigma}\!\mid\!m\rangle \langle m\!\mid\!a^\dagger_{\blb'\sigma}\!\mid\!i\rangle,
\end{equation}
where $Z=\sum_i\exp[-(E_i-\mu p_i)\beta]$ is the partition function, $\beta=1/T$, and $p_i$ is the number of electrons in the $i$-th state. For $T\rightarrow0$, this expression acquires the form
\begin{eqnarray}\label{T0}
C^{(1)}_{\blb'\blb}(\omega)&=&\langle 0\!\mid\!a_{\blb\sigma}(\omega+{\rm i}\eta+\mu+E_0-H_{\bf L})^{-1}a^\dagger_{\blb'\sigma}\!\mid\!0\rangle\nonumber\\
&&+\langle 0\!\mid\!a^\dagger_{\blb'\sigma}(\omega+{\rm i}\eta+\mu-E_0+H_{\bf L})^{-1}a^\dagger_{\blb'\sigma}\!\mid\!0\rangle,
\end{eqnarray}
where $\!\mid\!0\rangle$ is the ground state of the cluster Hamiltonian ${\cal H}_{\bf L}$ (\ref{cluster}). The energy of this state is equal to $E_0-\mu p_0$ with $E_0=\langle0\!\mid\!H_{\bf L}\!\mid\!0\rangle$ and the number of electrons in it $p_0=\langle0\!\mid\!\sum_{\blb\sigma}a^\dagger_{\blb\sigma}a_{\blb\sigma}\!\mid\!0\rangle$.  Deriving Eq.~(\ref{T0}), we carried out the analytic continuation to the real frequency axis from the upper half-plane and introduced the artificial broadening $\eta$.

In CPT, the crystal Green's function is obtained from Eq.~(\ref{Larkin}) disregarding replicas, which appear due to the crystal partitioning into clusters \cite{Senechal00,Senechal02},
\begin{equation}\label{Green}
G({\bf k},\omega)=\frac{1}{n}\sum_{\blb'\blb}{\rm e}^{{\rm i}{\bf k}(\blb'-\blb)}G_{\blb'\blb}({\bf k},\omega),
\end{equation}
where {\bf k} is the wave vector in the entire Brillouin zone of the crystal. In deriving this expression, the periodicity of $t_{\blb'\blb}({\bf q},\omega)$ and $G_{\blb'\blb}({\bf q},\omega)$ with respect to basis vectors of the reduced Brillouin zone was taken into account. The dependence of the electron concentration $x$ on the chemical potential is given by the relation
\begin{equation}\label{x_mu}
x(\mu)=-\frac{2}{N_xN_y\pi}\int_{-\infty}^{0}\sum_{\bf k}\,{\rm Im}\,G({\bf k},\omega)\,{\rm d}\omega,
\end{equation}
where $N_x$ and $N_y$ are numbers of sites in the periodic crystal region in the respective directions.

Notice that $x$ (\ref{x_mu}) generally differs from the electron concentration in the cluster $p_0/n$ used for calculating ${\bf C}^{(1)}$ (\ref{T0}). It happens due to the integration of clusters into the crystal, as described by Eq.~(\ref{Larkin}), and the formation of the energy band from cluster states. The band possesses a dispersion. When the chemical potential is located inside the band, its states with energies above $\mu$ are not occupied. As a consequence, $x\leq p_0/n$.

With the variation of $\mu$, the energies $E_0-\mu p_0$ of states with different numbers of electrons $p_0$ become the lowest one successively. In other words, the number of electrons in the cluster ground state alters with $\mu$. The change of $\!\mid\!0\rangle$ happens when the energies $E_0-\mu p_0$ and $E'_0-\mu p'_0$ of the states equalize, which occurs at $\mu_c=(E_0-E'_0)/(p_0-p'_0)$. The abrupt replacement of the ground state leads to a sharp swap of Green’s function (\ref{Green}) and a discontinuity in the dependence $x(\mu)$. The origin of this discontinuity is the same as that leading to the regions of negative electron compressibility in SCDT calculations \cite{Sherman20}. Hence the system of strongly correlated electrons is unstable with respect to phase separation. Similar conclusions were made in some other approaches \cite{Qin,Vandelli}. Chemical potentials considered in this work are far from $\mu_c$. For chemical potentials used in Sec.~4, the half-filled cluster state is the lowest one.

Ideally, Green's function (\ref{Green}) calculated with the exact irreducible part from Eq.~(\ref{Larkin}) should not depend on the size and geometry of clusters. Therefore,
\begin{eqnarray}\label{twopart}
&&\frac{1}{n}\sum_{\blb'\blb}{\rm e}^{{\rm i}{\bf k}(\blb'-\blb)}[{\bf 1-K(k,\omega) t(k)}]^{-1}{\bf K(k,\omega)}\rule[-1mm]{0.2mm}{4mm}_{\,\blb'\blb}\nonumber\\
&&\quad\quad=\{[K({\bf k},\omega)]^{-1}-t({\bf k})\}^{-1}.
\end{eqnarray}
In the right-hand side of the above equation, $K({\bf k},\omega)$ is the SCDT irreducible part. It is calculated using the series expansion in powers of the entire kinetic term of the Hamiltonian~(\ref{Hubbard}). Respectively, $t({\bf k})$ is the Fourier transform of $t_{\bf ll'}$. Equation~(\ref{twopart}) can be applied for calculating $K({\bf k},\omega)$ from its cluster counterpart ${\bf K(k,\omega)}$. It is especially useful for the case $T=0$ and real frequencies, where SCDT meets with difficulties.  In CPT, ${\bf K({\bf k},\omega)}$ is approximated by the first cumulant ${\bf C}^{(1)}(\omega)$. In this case, the SCDT irreducible part obtained from Eq.~(\ref{twopart}) contains processes taken into account in the cumulant.

\section{Magnetic field inclusion}
With the constant homogeneous magnetic field ${\bf B}$ perpendicular to the crystal plane, the Hubbard Hamiltonian takes the form
\begin{eqnarray}\label{Peierls}
H_B&=&-t\sum_{\bf la\sigma}\exp{\bigg(-{\rm i}\frac{\nu'a_x}{a}\bkp {\bf l}\bigg)}a^\dagger_{\bf l-a,\sigma}a_{\bf l\sigma}+\frac{1}{2}g\mu_BB\sum_{\bf l\sigma}\sigma n_{\bf l\sigma}\nonumber\\
&&+\frac{U}{2}\sum_{\bf l\sigma}n_{\bf l\sigma}n_{\bf l,-\sigma},
\end{eqnarray}
where the Peierls exponential factor \cite{Peierls} $\exp{[{\rm i}(e/\hbar)\int_{{\bf l-a}}^{{\bf l}}{\bf A(r)}d{\bf r}]}$ in the first term in the right-hand side of the equation is written using the Landau gauge ${\bf A(l)}=-Bl_y{\bf x}$. Here {\bf A} is the vector potential, $e$, $\hbar$, and {\bf x} are the modulus of the electron charge, Planck constant, and the unit vector along the $x$ axis, respectively. We restrict ourselves to fields satisfying the condition
\begin{equation}\label{condition}
\frac{e}{\hbar}B=2\pi\frac{\nu'}{\nu},
\end{equation}
where $\nu'$ and $\nu$ are coprime integers. This condition does not impose a rigid constraint on the value of $B$ since with large enough $\nu$ and $\nu’$, it can be approximated with the required accuracy. In Eq.~(\ref{Peierls}), $a_x$ is the $x$ component of the vector ${\bf a}$ connecting neighboring sites, $\bkp=2\pi/(\nu a){\bf y}$ with the unit vector along the $y$ direction {\bf y}, $g\approx2$ and $\mu_B$ are the $g$-factor and Bohr magneton, respectively.

The Peierls description used in Eq.~(\ref{Peierls}) is valid for fields with the magnetic length $l_B=\sqrt{\hbar/eB}$ much larger than the spatial extent of the Wannier function \cite{Peierls,Brown}. For fields $B\sim50$~T used in the magnetic-oscillation experiments $l_B\sim40$~\AA, and this condition is fulfilled. These values of $l_B$ far exceed linear sizes of clusters used in CPT, $\sqrt{n}a\lesssim10$~\AA, where the number of cluster sites $n\leq12$ and $a\approx3.8$~\AA\ for YBa$_2$Cu$_3$O$_{6+\delta}$. Using the Hamiltonian (\ref{Peierls}), we assume that the external magnetic field is only weakly disturbed by internal currents \cite{Atkinson}.

For such fields, The Zeeman term -- the second addend in the right-hand side of Eq.~(\ref{Peierls}) -- is two-three orders of magnitude smaller than the Hubbard repulsion, the bandwidth of noninteracting electrons, and the superexchange constant. Therefore, in this consideration, we neglect this term.

Let us consider a cluster at the location {\bf L} in the CPT cluster grid. For definiteness, we take a 2$\times$3 cluster. In accordance with Eq.~(\ref{Peierls}), the kinetic-energy part of its Hamiltonian reads
\begin{eqnarray}\label{kinetic}
H_{\bf L}^{\rm kin}&=&-t\sum_\sigma\bigg\{\sum_{\lambda_y=a}^{3a}\bigg[\exp{\bigg(-{\rm i}\frac{a(L_y+\lambda_y)}{l_B^2}\bigg)}\nonumber\\
&&\quad\quad\times a^\dagger_{{\bf L}+a{\bf x}+\lambda_y{\bf y},\sigma} a_{{\bf L}+2a{\bf x}+\lambda_y{\bf y},\sigma}+{\rm H.c.}\bigg]\nonumber\\
&&+\sum_{\lambda_y=a}^{2a}\sum_{\lambda_x=a}^{2a}\bigg[a^\dagger_{{\bf L}+\lambda_x{\bf x}+\lambda_y{\bf y},\sigma} a_{{\bf L}+\lambda_x{\bf x}+(\lambda_y+a){\bf y},\sigma}+{\rm H.c.}\bigg]\bigg\},
\end{eqnarray}
where $\lambda_x$ and $\lambda_y$ are components of the cluster site vector $\blb$, and H.c. designates a term, which is the Hermitian conjugate to that in front of it. Due to the mentioned inequality $\lambda_x,\lambda_y\ll l_B$ one can neglect $\lambda_y$ in the exponent in Eq.~(\ref{kinetic}). In such obtained Hamiltonian, the following gauge transformation can be performed:
\begin{eqnarray}
&&\tilde{a}^\dagger_{{\bf L}+a{\bf x}+\lambda_y{\bf y},\sigma}={\rm e}^{-{\rm i}aL_y/l_B^2} a^\dagger_{{\bf L}+a{\bf x}+\lambda_y{\bf y},\sigma},\nonumber\\[-1.5ex]
&&\label{gauge}\\[-1.5ex]
&&\tilde{a}^\dagger_{{\bf L}+2a{\bf x}+\lambda_y{\bf y},\sigma}=a^\dagger_{{\bf L}+2a{\bf x}+\lambda_y{\bf y},\sigma}.\nonumber
\end{eqnarray}
In terms of these operators with tildes, the cluster Hamiltonian for $B\neq0$ takes the form of that at $B=0$. That is, the ground state $\mid\!\!0\rangle_B$ and the cumulant ${\bf C}^{(1)}_B$ of the cluster in the magnetic field can be obtained from zero-field $\mid\!0\rangle$ and ${\bf C}^{(1)}$ from the previous section by substituting electron creation and annihilation operators by those with tildes (\ref{gauge}). Hence
\begin{eqnarray}\label{C_B}
&&C^{(1)}_B({\bf L}+s'a{\bf x}+\lambda_y,{\bf L}+sa{\bf x}+\lambda_y{\bf y})\nonumber\\
&&\quad={\rm e}^{{\rm i}(s'-s)aL_y/l_B^2}C^{(1)}({\bf L}+s'a{\bf x}+\lambda_y,{\bf L}+sa{\bf x}+\lambda_y{\bf y}),
\end{eqnarray}
where $s',s=1,2$. The exponent in the right-hand side contains the Peierls phase collected by an electron in its passage through the cluster. Similar relations between cumulants for $B\neq0$ and $B=0$ can be obtained for other CPT clusters.

The result (\ref{C_B}) points that, until the CPT with cluster sizes $\sqrt{n}a\ll l_B$ is a valid approximation, the actions of electron correlations and the magnetic field are fractionalized – the cumulant consists of a product of terms, one of which is connected with the Peierls phase and is not influenced by correlations, while the other describes correlations in the absence of the field. Due to the similarity between the SCDT and CPT, we can suppose that the zero-field value of $K$ obtained from Eq.~(\ref{twopart}) can be used for calculating the $B\neq0$ Green’s function also.

The transition to the SCDT description is necessary because the magnetic field reduces the translation symmetry of the problem. Indeed, as follows from Eq.~(\ref{Peierls}), the Hamiltonian is invariant for translations by a lattice period along the $x$ axis and by $\nu$ periods along the $y$ axis. The artificial crystal partitioning of the CPT and the translation periodicity in the magnetic field are, in general, incommensurate. It significantly complicates the consideration in the framework of CPT. The SCDT description with the zero-temperature, real-frequency irreducible part from Eq.~(\ref{twopart}) includes the information from CPT and is free of this difficulty.

For the mentioned translation symmetry of the Hamiltonian (\ref{Peierls}), it is convenient to define the spatial Fourier transformation as follows:
\begin{eqnarray}\label{Fourier}
a^\dagger_{{\bf q}'m\sigma}&=&\frac{1}{\sqrt{N_xN'_y\nu}}\sum_{\bf L'\blb'}{\rm e}^{{\rm i}({\bf q'}+m\bkp)({\bf L'+\blb'})}a^\dagger_{\bf L'+\blb',\sigma}\nonumber\\
&=&\frac{1}{\sqrt{N_xN'_y\nu}}\sum_{\bf L'\blb'}{\rm e}^{{\rm i}{\bf q'}({\bf L'+\blb'})+{\rm i}m\bkp\blb'}a^\dagger_{\bf L'+\blb',\sigma}.
\end{eqnarray}
Here ${\bf L'}$ labels the magnetic $1\times\nu$ clusters, $\blb'=sa{\bf y}$, $s=0,\ldots\nu-1$ is the site coordinate inside this cluster, $N'_y\nu=N_y$, ${\bf q'}$ is the wave vector in the reduced Brillouin zone with $N_x$ points in the $x$ direction and $N'_y$ points in the $y$ direction, and $m=0,\ldots\nu-1$. Notice that the electron movement along the $y$ direction is characterized by two parameters -- $q'_y$ and $m$.

In SCDT, we use the kinetic term of the Hamiltonian (\ref{Peierls}) as a perturbation and come to an equation similar to Eq.~(\ref{Larkin}). However, now matrix indices are variables $m$, and the hopping integral contains the Peierls factor,
\begin{equation}\label{Larkin_B}
{\bf G}({\bf q}',\omega)=[{\bf 1}-{\bf K}'({\bf q}',\omega){\bf T(q')}]^{-1}{\bf K}'({\bf q}',\omega),
\end{equation}
with $K'_{m'm}({\bf q}',\omega)=K(q'_x,q'_y+2\pi m/(\nu a))\delta_{m'm}$, where the irreducible part $K$ is derived from Eq.~(\ref{twopart}) with ${\bf K}={\bf C}^{(1)}$ taken from the CPT calculations, and {\bf T} is the Fourier transform of the hopping integral with the Peierls factor,
\begin{eqnarray}\label{T}
T_{m'm}({\bf q}')&=&-t\Bigg[{\rm e}^{{\rm i}aq'_x}\delta_{m',m+\nu'}+{\rm e}^{-{\rm i}aq'_x}\delta_{m',m-\nu'}\nonumber\\
&&+2\cos\left(aq'_y+m\frac{2\pi}{\nu}\right)\delta_{m'm}\Bigg].
\end{eqnarray}

\section{Magnetic oscillations}
\begin{figure}[t]
\centerline{\resizebox{0.8\columnwidth}{!}{\includegraphics{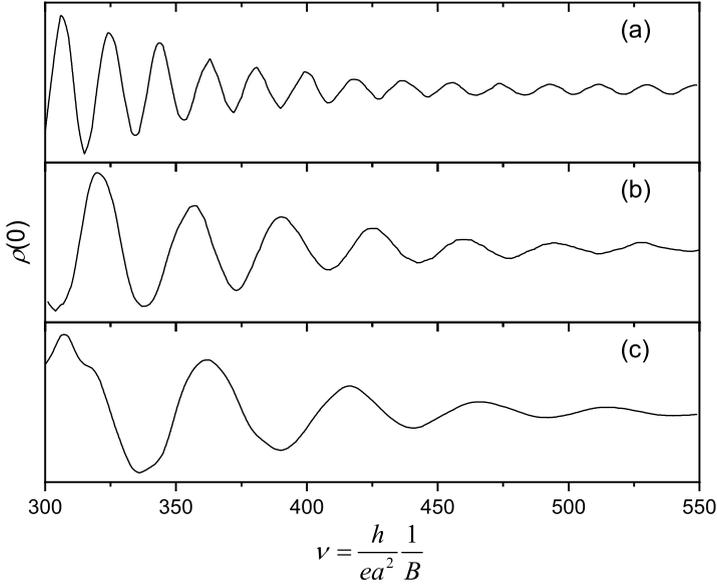}}}
\caption{The density of electron states at the Fermi level as a function of $\nu=h/(ea^2B)$. $U=8t$, $\eta=0.01t$, $\nu'=1$, $x\approx0.949$ (a), 0.955 (b), and 0.961 (c).} \label{Fig2}
\end{figure}
Quantum oscillation experiments measure variations of different derivatives of the thermodynamic potential $\Omega(T,V,\mu,B)$ with the magnetic field \cite{Shoenberg}. Here $V$ is the volume. Since the electron DOS on the Fermi level $\rho(0)$ is the second derivative of $\Omega$ with respect to $\mu$, the change of $\rho(0)$ with $B$ provides insight into the dependence $\Omega(B)$. The relation between the DOS and Green's function (\ref{Larkin_B}) reads
\begin{equation}\label{rho}
\rho(\omega)=-\frac{2}{N_xN_y\pi}\sum_{{\bf q}'m}{\rm Im}\,G_{mm}({\bf q}',\omega).
\end{equation}

In accordance with the discussion of the previous sections, for calculating $\rho(0)$ (\ref{rho}), we found the first cumulant ${\bf C}^{(1)}$ (\ref{T0}) for a 12-site cluster, derived the irreducible part $K$ from the relation~(\ref{twopart}), and used it in Eq.~(\ref{Larkin_B}).

Results of these calculations are shown in Fig.~\ref{Fig2} for different doping. In all cases, the dependence $\rho(0)$ on $1/B$ is close to harmonic. The values $\nu\gg1$ correspond to experimentally used fields. In this case, we can set $\nu'=1$ and obtain smooth curves. The quantum oscillation frequency $F=1/\Delta(1/B)$ is equal to 1510, 777, and 504~T for dependencies (a), (b), and (c), respectively. Here $\Delta(1/B)$ is the oscillation period in the curves. The latter frequency is comparable to that observed experimentally \cite{Sebastian}. It is more than an order of magnitude smaller than frequencies of magnetic oscillations in conventional metals, which points to a small Fermi surface. Its size varies considerably in the small range of electron concentrations. Notice that all three curves in Fig.~\ref{Fig2} were obtained with the same cumulant ${\bf C}^{(1)}$ corresponding to a half-filled cluster. In the considered range of $\mu$, this cluster state is the lowest one.

\begin{figure}[t]
\centerline{\resizebox{0.8\columnwidth}{!}{\includegraphics{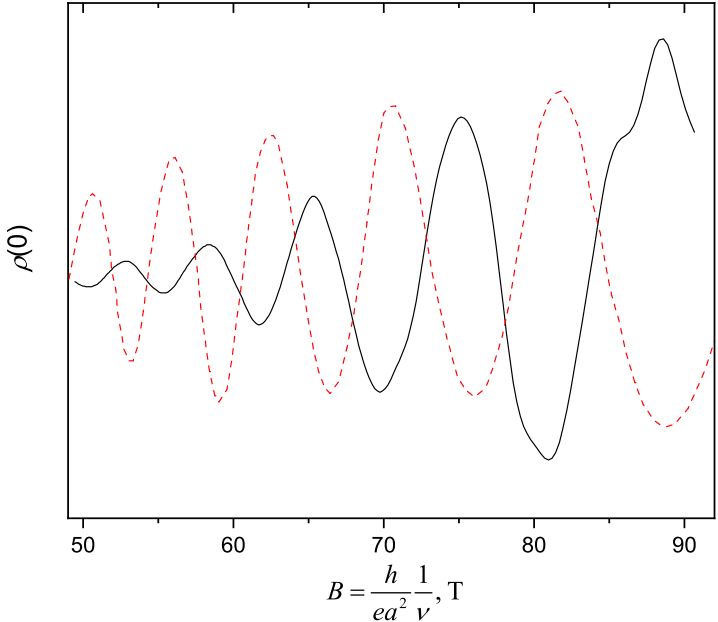}}}
\caption{The comparison of magnetic oscillations observed experimentally in YBa$_2$Cu$_3$O$_{6.56}$ (Fig.~2(a) in \protect\cite{Sebastian}, red dashed curve) and that shown in Fig.~\protect\ref{Fig2}(c) (black solid curve).} \label{Fig3}
\end{figure}
As mentioned, for parameters of Fig.~\ref{Fig2}(c), the quantum oscillation frequency is close to that observed in lightly doped YBa$_2$Cu$_3$O$_{6+\delta}$. Figure~\ref{Fig3} compares the experimental and calculated curves. They are displaced in phase by nearly half a period. However, their frequencies are close.

\begin{figure}[t]
\centerline{\resizebox{0.99\columnwidth}{!}{\includegraphics{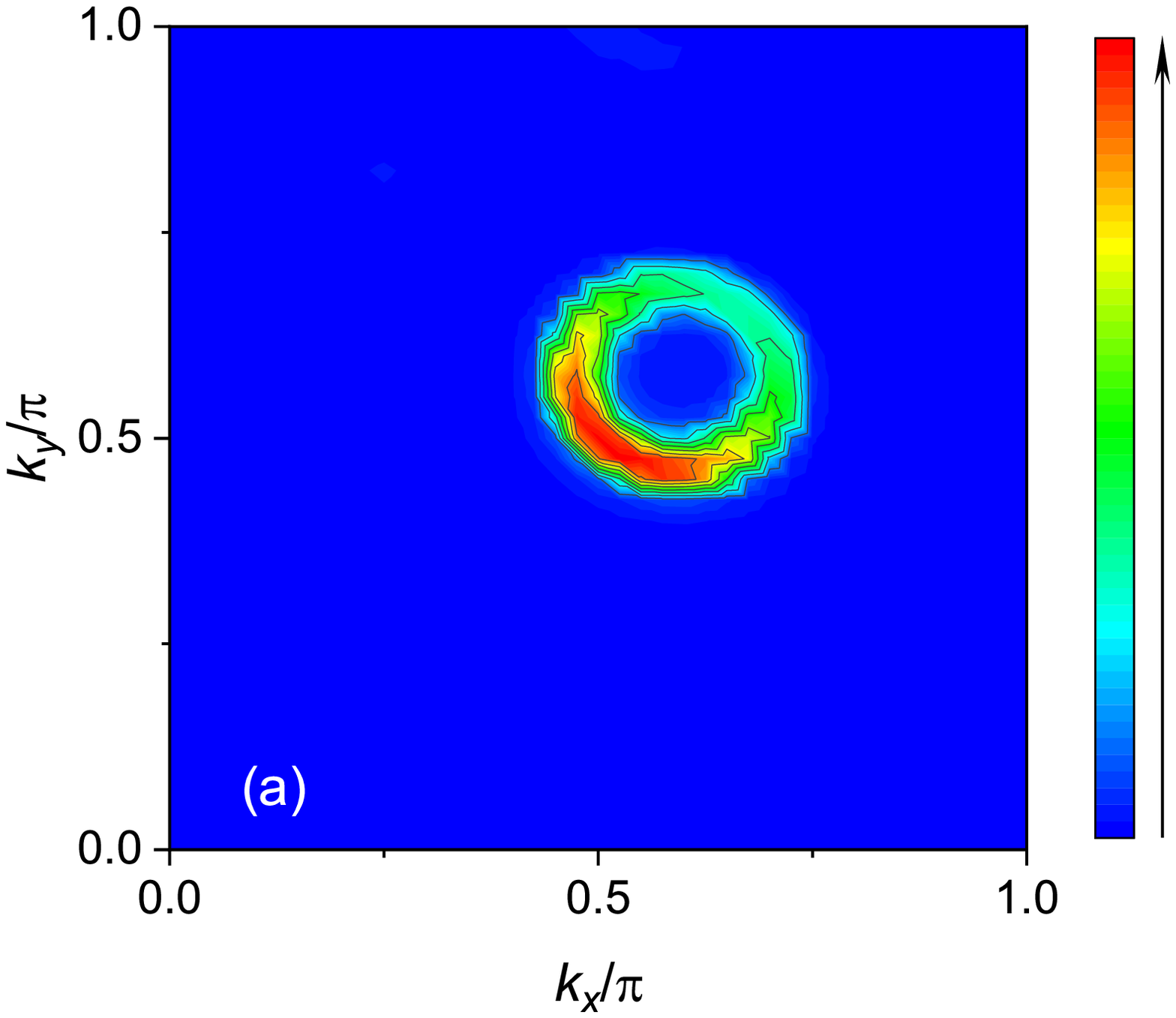}\hspace{2em}\includegraphics{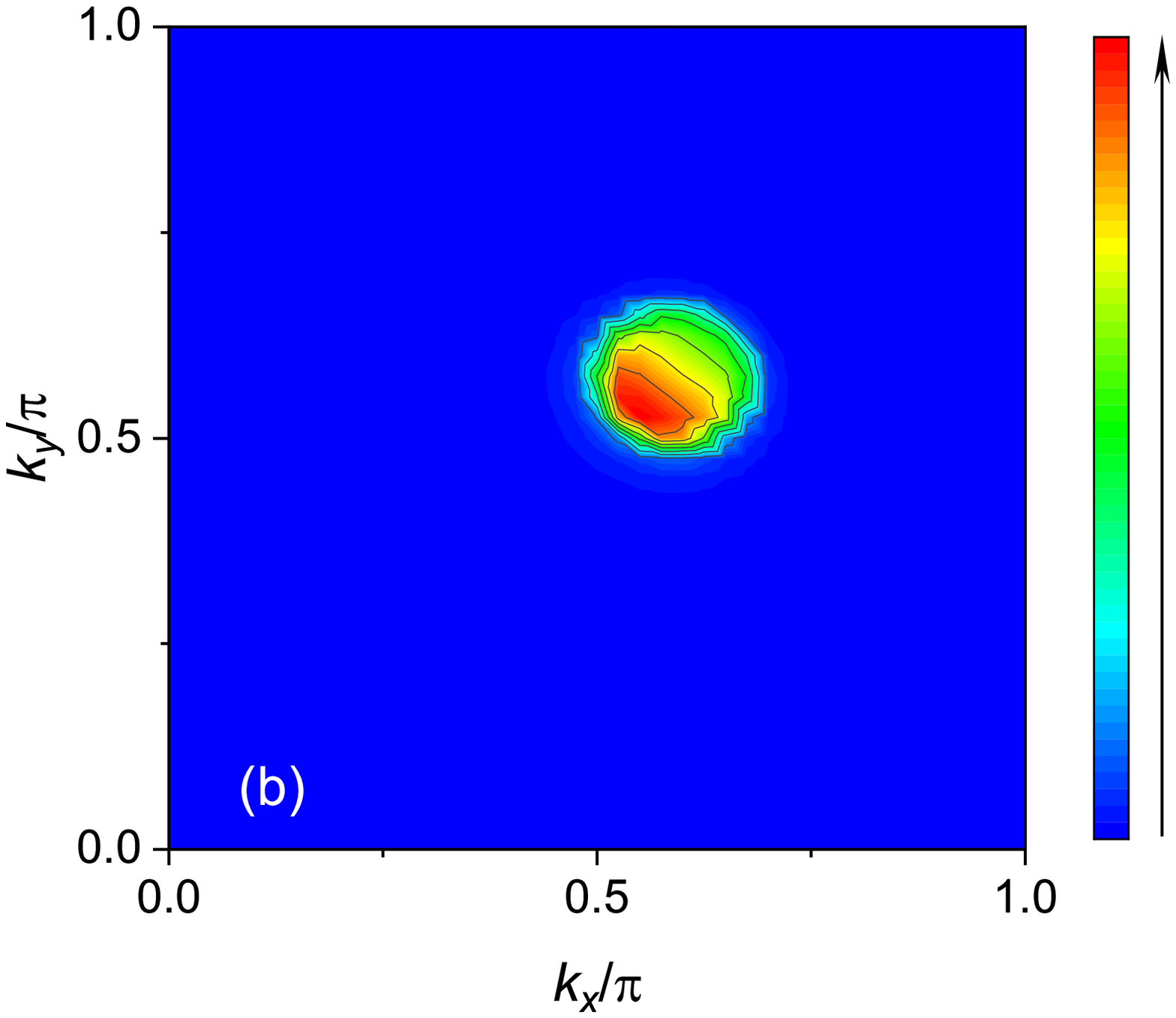}}}
\caption{Fermi surfaces in the first quadrant of the Brillouin zone. Parameters are the same as in figure~\protect\ref{Fig2}(a) (panel (a)) and \protect\ref{Fig2}(b) (panel (b)).} \label{Fig4}
\end{figure}
Fermi surfaces for parameters of panels (a) and (b) in Fig.~\ref{Fig2} are shown in Fig~\ref{Fig4}. The surface corresponding to panel (c) appears as a reduced version of the image in Fig.~\ref{Fig4}(b). Data for these pictures were obtained by integrating the spectral function $A({\bf k},\omega)=-\pi^{-1}{\rm Im}\,G({\bf k},\omega)$ over the frequency range of the width $0.1t$ below the Fermi level. For small deviations from half-filling, the Fermi surface is located near nodal points. It consists of a Fermi arc supplemented with a less intensive part. Together they form a nearly circular region that occupies several percent of the Brillouin zone. A small area $S$ of the region explains low oscillation frequencies $F$ in Figs.~\ref{Fig2} and \ref{Fig3}, since $F\propto S$ \cite{Shoenberg}. A similar picture was suggested in Ref.~\cite{Sebastian} from the analysis of experimental results.

\section{Concluding remarks}
Using the cluster perturbation theory and strong coupling diagram technique, we investigated the system of electrons described by the two-dimensional Hubbard model and resided in a perpendicular to the crystal plane homogeneous magnetic field. For magnetic fields $B\sim50$~T and small deviations from half-filling $1-x\approx0.04$, we found magnetic oscillations of the electron density of states at the Fermi level with frequencies $F\sim500-1000$~T. Such frequencies are an order of magnitude smaller than those observed in conventional metals and close to the oscillations seen in lightly doped yttrium cuprates. As follows from our calculations, the reason for these low frequencies is small closed Fermi surfaces near nodal points of the Brillouin zone. The surfaces are formed by Fermi arcs and less intensive sections, which together produce nearly circular contours. Such Fermi surfaces were presumed from the analyses of experimental results. However, photoemission data on less intensive parts of the surfaces were contradictory. Calculated low oscillation frequencies are connected with small arias of the Fermi surfaces, which occupy only a few percent of the Brillouin zone.

\backmatter

\section*{Declarations}
Conflict of interest: The author declares that he has no conflict of interest. There are no other applicable declarations.

\end{document}